\documentclass[a4paper,12pt]{article}

\usepackage{latexsym}
\usepackage{amsfonts} 
\usepackage{ascmac}
\usepackage[mathscr]{euscript}
\usepackage{amsmath,mathrsfs,bm,amssymb,color}

\newcommand{\D}{\mathrm{dom}}

\newcommand{\la}{\langle}
\newcommand{\ra}{\rangle}
\newcommand{\Tr}{\mathrm{Tr}}

\newcommand{\BbbR}{\mathbb{R}}
\newcommand{\BbbN}{\mathbb{N}}

\newcommand{\BbbC}{\mathbb{C}}

\newcommand{\vphi}{\varphi}

\newcommand{\no}{\nonumber \\}

\newcommand{\Ue}{U_{\mathrm{eff}}}
\newcommand{\ef}{\mathrm{eff}}

\newcommand{\mb}{\mathbf}
\newcommand{\bfa}{{\boldsymbol a}}
\newcommand{\bfb}{{\boldsymbol b}}

\newcommand{\hh}{H_{\mathrm{H}}}

\begin{document}

\newtheorem{define}{Definition}
\newtheorem{Thm}[define]{Theorem}
\newtheorem{Prop}[define]{Proposition}
\newtheorem{lemm}[define]{Lemma}
\newtheorem{rem}[define]{Remark}
\newtheorem{assum}{Condition}
\newtheorem{example}{Example}
\newtheorem{coro}[define]{Corollary}

\title{Ground state properties of the Holstein-Hubbard model}
\author{
Tadahiro Miyao\\
 Department of Mathematics,
 Hokkaido University,\\
Sapporo 060-0810, Japan\\
E-mail:
 miyao@math.sci.hokudai.ac.jp
}

\date{\empty}

\maketitle

\begin{abstract}
We study the ground state properties of the Holstein-Hubbard model on some bipartite lattices 
at half-filling; 
The    ground state is   proved   to   exhibit ferrimagnetism whenever the electron-phonon
interaction is not so strong.
  In addition, the antiferromagnetic long range order is shown to exist in
 the ground state.  In contrast to
this, we prove  the absence of  the  long range charge order. 
\begin{flushleft}
\end{flushleft} 
\end{abstract} 



\section{Introduction and results}

To explain ferromagnetism from the Hubbard model is  known as 
a challenging problem. Since  the discovery of the  Nagaoka-Thouless
ferromagnetism
 \cite{NT, Thouless}, 
there have been significant developments in this field:
 The ground state  of the Hubbard model on some bipartite lattices at
 half-filling is shown  to exhibits ferrimagnetism  by Lieb \cite{Lieb}; 
 Mielke \cite{Mielke,Mielke2, Mielke3, Mielke4} and Tasaki \cite{Tasaki,
 Tasaki2, Tasaki3} constructed rigorous examples of ferromagnetic ground states 
in certain  Hubbard models. However, the origin of ferromagnetism is
still incompletely understood.

In the presence of electron-electron Coulomb and electron-phonon interaction,
correlated electron systems provide an attractive field of study.
The Holstein-Hubbard model is a simple model describing the interplay of 
electron-electron and electron-phonon interactions.
Despite its importance, rigorous studies   of magnetic properties  of  
the Holstein-Hubbard model are   rare; see, e.g. \cite{FL}.  
Recently,  Miyao proved that the ground state of the
Holstein-Hubbard model
 on some bipartite lattices at half-filling is unique  whenever the electron-phonon
 interaction is not so strong \cite{Miyao}.

 In the present paper, we prove that
the unique ground state exhibits ferrimagnetism (Theorem \ref{HH})
 as an important consequence of \cite{Miyao}.
As far as we know, this is a first rigorous example of ferrimagnetism in the 
Holstein-Hubbard model.
The idea of our proof is to extend Lieb's mathod in \cite{Lieb}.
In addition, we prove the  
existence of antiferromagnetic long range order (Theorem \ref{LRO}) and 
  absence of the long range charge order (Theorem \ref{Ab}) in the 
ground state.

The Hamiltonian of the   Holstein--Hubbard model on a finite lattice $\Lambda$ is given by 
\begin{align}
H_{\mathrm{HH}}=&\sum_{x,y\in \Lambda}
\sum_{
\sigma\in \{\uparrow, \downarrow\}}t_{xy}
 c_{x\sigma}^*c_{y\sigma}+\sum_{x, y\in
 \Lambda}\frac{U_{xy}}{2}(n_{x}-1)(n_y-1)\no
&+\sum_{x, y\in \Lambda}
g_{xy}n_{x}(b_y^*+b_y)+\sum_{x\in \Lambda} \omega b_x^*b_x, \label{ExtendedHH}
\end{align}  
where $c_{x\sigma}$  is the electron annihilation operator at site  $x$ and $b_x$ is
the phonon annihilation operator at site $x$. These operators satisfy 
the following relations:
\begin{align}
\{c_{x\sigma}, c_{x'\sigma'}^*\}=\delta_{\sigma\sigma'}\delta_{xx'},\ 
[b_x, b_{x'}^*]=\delta_{xx'},
\end{align} 
 where
 $\delta_{xy}$ is the Kronecker delta.
 $n_{x}$ is the fermionic number
 operator at site $x\in \Lambda$ defined by 
$
n_x=\sum_{\sigma\in \{\uparrow, \downarrow\}} n_{x\sigma},\ 
 n_{x\sigma}=c_{x\sigma}^* c_{x\sigma}
$.
$t_{xy}$ is the hopping matrix element, $U_{xy}$ is the energy of the
Coulomb interaction, and $g_{xy}$ is the strength of the electron-phonon
interaction.  We assume that 
  $\{g_{xy}\}, \{t_{xy}\}$ and $\{U_{xy}\}$ are real symmetric
 $|\Lambda|\times |\Lambda|$ matrices.\footnote{
Let $M=\{M_{xy}\}$  be a $|\Lambda|\times|\Lambda|$ matrix. $M$ is
 called a {\it real symmetric matrix}  if $M_{xy}$ is real and $M_{xy}=M_{yx}$ for
 all $x,y\in \Lambda$.
 }
The phonons are assumed to be dispersionless with energy $\omega>0$.

$H_{\mathrm{HH}}$ acts on the Hilbert space $\mathfrak{E} \otimes \mathfrak{P}$, where 
$\mathfrak{E}=\bigoplus_{n\ge 0} \wedge^n (\ell^2(\Lambda) \oplus
\ell^2(\Lambda))$, the fermionic Fock space and $\mathfrak{P}=\bigoplus_{n\ge 0}
\otimes_{\mathrm{s}} ^n \ell^2(\Lambda)$, the bosonic Fock space.
Here, $\wedge^n (\ell^2(\Lambda) \oplus \ell^2(\Lambda))$ indicates the
$n$-fold antisymmetric tensor product of $\ell^2(\Lambda) \oplus
\ell^2(\Lambda)$, while $\otimes_{\mathrm{s}} ^n \ell^2(\Lambda)$
indicates the $n$-fold symmetric tensor product of $\ell^2(\Lambda)$. 

$H_{\mathrm{HH}}$ is self-adjoint on $\D(N_{\mathrm{b}})$ and bounded from below, 
where $N_{\mathrm{b}}=\sum_{x\in \Lambda} b_x^*b_x$ and  $\D(A)$ is the
domain of the linear operator $A$.
\medskip

\begin{rem}{\rm
At a first glance, it appears that  the Coulomb interaction term in (1) is
not standard;   however,  our Coulomb interaction coincides with  the standard one when 
$
\sum_{x\in \Lambda}U_{xy}$
	is a constant 
	independent of $y$;  in this case, the  Coulomb interaction in (1) becomes
\begin{align}
 \frac{1}{2}\sum_{x, y\in \Lambda}U_{xy}(n_x-1)(n_y-1)=\sum_{x\in \Lambda}U_{xx}
	n_{x\uparrow}n_{x\downarrow}+\frac{1}{2}\sum_{x\neq
	y}U_{xy}n_xn_y
+Const.
\end{align} 
 for every electron filling. 
A typical example satisfying the assumption  about $U_{xy}$ is 
the case where  $U_{xy}=U_0 \delta_{xy}$, see also Remark \ref{Perio}. $\diamondsuit$
}
\end{rem}

We say that there is a {\it bond} between $x$ and $y$ if
	$t_{xy}\neq 0$.
We impose the following conditions on $\Lambda$:
\begin{flushleft}
{\bf (A. 1)} $\Lambda$ is connected, namely, there is a connected path of
bonds between every pairs of sites.\footnote{More precisely, for any $x, y\in \Lambda$, there exist $x_1,
 \dots, x_n\in \Lambda$ such that $x_1=x,\ x_n=y$ and $
t_{x_1x_2}t_{x_2x_3}\cdots t_{x_{n-1} x_n} \neq 0
$.}
\medskip

{\bf (A. 2)} $\Lambda$ is bipartite, namely, $\Lambda$ can be divided
into two disjoint sites $A$ and $B$ such that $t_{xy}=0$
 whenever $x, y\in A$ or $x, y\in B$.
\end{flushleft}

As to the electron-phonon interaction, we assume the following condition:
\begin{flushleft}
{\bf (A. 3)}
  $\displaystyle 
\sum_{x\in \Lambda} g_{xy}$ is  a constant independent of $y\in \Lambda$ .
\end{flushleft}
 
\begin{rem}
{\rm 
\begin{itemize}
\item[(i)] A typical  example satisfying {\bf (A. 3)} is
 $g_{xy}=g_0\delta_{xy}$,
 see also Remark \ref{Perio}.
\item[(ii)] Let us consider a linear chain of $2L$ atoms with periodic boundary
 conditions.  We set $\Lambda=\{x_j\}_{j=1}^{2L}$.
 Assume that $|x_j-x_{j+1}|=\mathrm{constant}$ for all $j$, where $x_{2L+1}=x_1$.
If $g_{xy}$ is a function
 of $|x-y|$, i.e., $g_{xy}=f(|x-y|)$, then {\bf (A. 3)} is satisfied. Similarly,
 if $\Lambda$ has a symmetric structure, like $\mathrm{C}_{60}$
 fullerene, then {\bf (A. 3)} is fulfilled.  $\diamondsuit$ 
 
 \end{itemize}}
 
\end{rem}

Let $N_{\mathrm{el}}$ be the electron number operator given by 
$
N_{\mathrm{el}}=\sum_{x\in \Lambda} n_x
$. Trivially, we have 
$\mathrm{spec}(N_{\mathrm{el}})=\{0, 1, \dots, 2|\Lambda|\}$, where 
$\mathrm{spec}(N_{\mathrm{el}})$ indicates the spectrum of $N_{\mathrm{el}}$.
We can decompose the Hilbert space $\mathfrak{E} \otimes \mathfrak{P}$
as 
\begin{align}
\mathfrak{E} \otimes \mathfrak{P}=\bigoplus_{n=0}^{2|\Lambda|}
 \mathfrak{E}_n\otimes \mathfrak{P},
\end{align} 
where $\mathfrak{E}_n=\wedge^n\big( \ell^2(\Lambda) \oplus \ell^2(\Lambda)\big)$,
 the $n$-electron subspace. Of course, $
\mathfrak{E}_n=\ker(N_{\mathrm{el}}-n)$.
The number of electron is conserved, i.e., $H_{\mathrm{HH}}$ commutes
 with $N_{\mathrm{el}}$. Hence, $H_{\mathrm{HH}}$ can be decomposed as 
\begin{align}
H_{\mathrm{HH}} =\bigoplus_{n=0}^{2|\Lambda|} H_{\mathrm{HH}, n},\quad
H_{\mathrm{HH},n}=H_{\mathrm{HH}} \restriction \mathfrak{E}_n\otimes
 \mathfrak{P},
\end{align} 
where $H_{\mathrm{HH}} \restriction \mathfrak{E}_n\otimes \mathfrak{P}$
 is the restriction of $H_{\mathrm{HH}}$ on $\mathfrak{E}_n\otimes
 \mathfrak{P}$.
Because we are interested in the half-filled case, we will study the
Hamiltonian 
\begin{align}
H:=H_{\mathrm{HH}, n=|\Lambda|}.
\end{align}

Let $S_x^{(+)}=c_{x\uparrow}^*c_{x\downarrow}$ and let
$S_x^{(-)}=(S_x^{(+)})^*$.  The spin operators are defined by
\begin{align}
S^{(3)}=\frac{1}{2}\sum_{x\in \Lambda}(n_{x\uparrow}-n_{x\downarrow}),
\ S^{(+)}=\sum_{x\in \Lambda}S_x^{(+)},\ S^{(-)}=\sum_{x\in
\Lambda}S_x^{(-)}.
\end{align} 
The total spin operator is defined  by 
\begin{align}
S_{\mathrm{tot}}^2=(S^{(3)})^2+\frac{1}{2}S^{(+)}S^{(-)}+
\frac{1}{2}S^{(-)}S^{(+)}
\end{align} 
with eigenvalues $S(S+1)$.
Let $\vphi$ be a vector in $\mathfrak{E}_{n=|\Lambda|} \otimes
\mathfrak{P}$.
If $\vphi$ is an eigenvector of $S_{\mathrm{tot}}^2$
with $S_{\mathrm{tot}}^2\vphi=S(S+1)\vphi$, then we say that $\vphi$ has
total spin $S$.
Main purpose in the present paper is to study the total spin $S$
 for the ground states.

To state our results, we introduce the  effective Coulomb interaction  by 
\begin{align}
U_{\ef, xy}=U_{xy}-\frac{2}{\omega}\sum_{z\in \Lambda}g_{xz}g_{yz}.
\end{align}

\begin{Thm}\label{HH}
Assume that $|\Lambda|$ is even. Assume {\bf (A. 1)}---{\bf (A. 3)}.
Assume that  $\{U_{\mathrm{eff}, xy}\}$ is positive definite.\footnote{
A matrix $\{M_{xy}\}$ is  called {\it  positive definite} if 
$
\sum_{x, y\in \Lambda}\overline{\xi}_x \xi_y M_{xy}> 0
$ (strict inequality)
holds for 
 all $\{\xi_x\}_{x\in \Lambda}\in \BbbC^{|\Lambda|}\backslash\{{\bf
	0}\}$. 
}
Then the ground state of $H$  has total spin $S=\frac{1}{2}\big||B|-|A|\big|$
 and 
is unique apart from the trivial 
$(2S+1)$-degeneracy.
\end{Thm}
\begin{rem}{\rm 
\begin{itemize}
\item[(i)] In general, the positive definitness of $\{U_{\mathrm{eff}, xy}\}$ implies
 that the electron-phonon interaction is not so strong. To see this,    
consider the case where  $U_{xy}=U_0\delta_{xy}$ and $g_{xy}=g_0\delta_{xy}$.
In this case, $H$ becomes the standard Holstein-Hubbard model.
 $\{U_{\ef, xy}\}$
 is positive definite if and only if $|g_0|<\sqrt{\omega
 U_0/2}$, namely, the electron-phonon interaction is not so strong.
\item[(ii)]  Theorem \ref{HH} claims that Lieb's ferrimagnetism (Theorem \ref{Hubbard})
is stable whenever the electron-phonon interaction is not so strong. 

\item[(iii)] Recently, Nagaoka's theorem in the Hubbard model is extended to the Holstein-Hubbard model \cite{Miyao2}.
Theorem \ref{HH} is consistent with this result. 
\item[(iv)] In \cite{Miyao3}, Theorem \ref{HH} is examined from a view point of  universality.
$\diamondsuit$
\end{itemize} 
}
\end{rem}

\begin{rem}\label{Perio}
{\rm 
Let $\mathscr{P}$ be a Bravais lattice with  the  set of primitive vectors 
$\{\bfa_1,\dots, \bfa_d\}$ ($d=2, 3$).
If $\Lambda$ is a subset  of $\mathscr{P}$, then the positive
 definitness of $\{U_{\mathrm{eff}, xy}\}$ can be expressed as follows:
Let   $\{\bfb_1,\dots,  \bfb_d\}$ be the  set of primitive vectors of the reciprocal
 lattice of $\mathscr{P}$, i.e., $\bfa_i\cdot \bfb_j=2\pi \delta_{ij}$. 
We set 
$
\Lambda=\Big\{\sum_{j=1}^d n_j\bfa_j\, \Big|\, n_j=-L+1,  \dots, L
\Big\}
$ 
 and $
\Lambda^*=\Big\{\sum_{j=1}^d \ell_j\bfb_j/L\, \Big|\ell_j=-L+1, \dots, L \Big\}.$
 Suppose that $g_{xy}$ and $U_{xy}$ are given by 
\begin{align}
g_{xy}=\frac{1}{ |\Lambda|}\sum_{k \in \Lambda^*} G(k) e^{i k\cdot
 (x-y)},
\ \ \ 
U_{xy}=\frac{1}{ |\Lambda|}\sum_{k \in \Lambda^*} U(k) e^{i k\cdot
 (x-y)},
\end{align} 
where $G(k)$ and $U(k)$ are real-valued  continuous functions on $T_d
=\Big\{\sum_{j=1}^d \theta_j \bfb_j\, 
|-1 \le \theta_j \le 1 \Big\}
$ with $G(-k)=G(k)$ and $U(-k)=U(k)$.
Since $
\sum_{x\in \Lambda} g_{xy}= G(0)
$ for all $y\in \Lambda$, {\bf (A. 3)} is satisfied.  In this case, we obtain
\begin{align}
U_{\mathrm{eff}, xy}=\frac{1}{ |\Lambda|}\sum_{k \in \Lambda^*} 
\Bigg\{
U(k)-\frac{2}{\omega} G(k)^2
\Bigg\}
 e^{i k\cdot (x-y)}.
\end{align} 
If $
U(k)>\frac{2}{\omega}G(k)^2
$ for all $k\in T_d$, then $U_{\mathrm{eff}, xy}$ is
 positive definite for  all $L\in \BbbN$. It is noteworthy that this condition  
 is uniform in the size.
 Similarly, we can handle $g_{xy}$ and $U_{xy}$ on more complicated lattices  (e.g.,  the Lieb lattice etc.).
   $\diamondsuit$
}
\end{rem}

Let 
\begin{align}
\hat{S}_0^{(+)}=|\Lambda|^{-1/2}\sum_{x\in \Lambda} S_x^{(+)}, \ \ \ \ 
\hat{S}^{(+)}_Q=|\Lambda|^{-1/2}\sum_{x\in \Lambda} \gamma(x)
S_x^{(+)},
\end{align} 
where $\gamma(x)=1$ if $x\in A$, $\gamma(x)=-1$ if $x\in B$.
 The correlation functions are  given by 
\begin{align}
m(k)=\Big\la \hat{S}_k^{(+)}\big(\hat{S}_{k}^{(+)}\big)^*\Big\ra
\end{align} 
 for  $k=0$ or $Q$, where $\la\cdot
\ra$ is the ground state expectation.
\begin{Thm}\label{LRO}
Assume that $|\Lambda|$ is even. Assume {\bf (A. 1)}---{\bf (A. 3)}.
Assume that  $\{U_{\mathrm{eff}, xy}\}$ is positive definite.
If $\big||A|-|B|\big|=const. |\Lambda|$, then 
\begin{align}
m(Q) \ge m(0)=O(|\Lambda|).
\end{align} 
Thus,  the antiferromagnetic and ferrimagnetic long range order coexist 
in the ground state.
\end{Thm}

Finally, we present  a theorem on the charge susceptibility.
Suppose that  that $\Lambda,\  g_{xy}$ and $U_{xy}$ 
 are given in Remark \ref{Perio}.
Let $q_x=n_x-1$. The charge susceptibility (at $\beta =\infty$) with the
wave vector $k$ is given by 
\begin{align}
\chi_k=\big\la \hat{q}_k(H-E)^{-1} \hat{q}_{-k}\big\ra,
\end{align} 
where $\hat{q}_k=|\Lambda|^{-1/2} \sum_{x\in \Lambda} e^{-ik\cdot x}
q_x$
 and $E$ is the ground state energy of $H$.
\medskip

\begin{Thm}\label{Ab}
Assume that $|\Lambda|$ is even.
Assume that  $\{U_{\mathrm{eff}, xy}\}$ is positive semidefinite\footnote{
A matrix $\{M_{xy}\}$ is called {\it  positive semidefinite} if, for
 all $\{\xi_x\}_{x\in \Lambda}\in \BbbC^{|\Lambda|}$, 
$
\sum_{x, y\in \Lambda}\overline{\xi}_x \xi_y M_{xy}\ge  0
$
holds.
},
that is, $U(k) \ge \frac{2}{\omega}G(k)^2$ for all $k\in T_d$. 
Then we  have
\begin{align}
\chi(k) \le \frac{1}{U_{\mathrm{eff}}(k)},
\end{align} 
where $U_{\mathrm{eff}}(k)=U(k)-\frac{2}{\omega} G(k)^2$.
 Thus, if 
there exists a constant $c_0>0$ such that 
$U_{\mathrm{eff}}(k)\ge c_0$ for all $k\in T_d$, then there is no   long range charge order.
\end{Thm}

\begin{rem}
{\rm
Theorems \ref{LRO} and  \ref{Ab} suggest   that  coexistence of  the ferrimagnetic and
	     charge long range  orders  would be  impossible.
	    For instance, consider  the model on the Lieb lattice with $U_{xy}=U_0\delta_{xy}$
	    and $g_{xy}=g_0\delta_{xy}$. 
	    Suppose that $|g_0| <\sqrt{\omega U_0/2}$.
	    By Theorem \ref{Ab}, we have
\begin{align}	    
\chi(k) \le (U_0-2g_0^2/\omega)^{-1},
\end{align} 
 which implies the absence of the 
	     long range charge order. On the other hand, Theorem \ref{LRO}
	    claims the coexistence of the  ferrimagnetic and antiferromagnetic long range  orders. $\diamondsuit$
}

\end{rem}

\section{Proofs}
\subsection{Preliminaries:
An extension of  Lieb's theorem}
We denote the spectrum of $S^{(3)}$ by $\mathrm{spec}(S^{(3)})$.
Remark that $
\mathrm{spec}(S^{(3)})=\{-|\Lambda|/2, -|\Lambda|/2+1, \dots, |\Lambda|/2\}
$. For each $M\in \mathrm{spec}(S^{(3)})$, 
we set 
\begin{align}
\mathcal{H}_M:=\big(\mathfrak{E}_{n=|\Lambda|} \otimes \mathfrak{P} \big) \cap
 \ker\big(S^{(3)}-M \big).
\end{align} 
 We call $
\mathcal{H}_M
$  the $S^{(3)}=M$ subspace.

The following theorem is a basic input in the present paper.
\medskip

\begin{Thm}{\rm \cite{Miyao}}\label{Uni}
Assume that $|\Lambda|$ is even. Assume {\bf (A. 1)}---{\bf (A. 3)}.
Assume that  $\{U_{\mathrm{eff}, xy}\}$ is positive definite.
 For each $M\in \{-|\Lambda|/2,
 -|\Lambda|/2+1,\dots, |\Lambda|/2\}$,
the ground state of $H$ is unique in each $S^{(3)}=M$ subspace.  Let $\vphi_M$ be the unique ground
 state of $H$ in the $S^{(3)}=M$ subspace. Then the following holds:
\begin{align}
\big\la \vphi_M| S_{x}^{(+)}S_{y}^{(-)}\vphi_M\big\ra
\begin{cases}
>0\ \ \mbox{if $x, y\in A$ or $x, y\in B$}\\
<0\ \ \mbox{otherwise}. \label{GSAnti2}
\end{cases}
\end{align} 
\end{Thm}
Remark that the proof of Theorem \ref{Uni}  is based on operator theoretic correlation inequalities.\footnote{
Whereas the subjects are  different, 
there are some similarities between  the ideas in \cite{Miyao} and \cite{MN} . }

From  Theorem \ref{Uni}, we can derive an extension of   Lieb's theorem \cite{Lieb}. Let $H_{\mathrm{H}}$
be the extended Hubbard model defined  by 
\begin{align}
H_{\mathrm{H}}=&\sum_{x, y\in \Lambda}
\sum_{
\sigma\in \{\uparrow, \downarrow\}}t_{xy}
 c_{x\sigma}^*c_{y\sigma}+\sum_{x, y\in
 \Lambda}\frac{U_{xy}}{2}(n_{x}-1)(n_y-1).
 \label{ExtendedH}
\end{align}

\begin{Thm}\label{Hubbard}
Assume that $|\Lambda|$ is even. Assume {\bf (A. 1)}---{\bf (A. 3)}.
Assume that  $\{U_{xy}\}$ is positive definite.
Then 
the ground state of $H_{\mathrm{H}}$ has total spin
 $S=\frac{1}{2}\big||B|-|A|\big|$ and is unique apart from the trivial 
$(2S+1)$-degeneracy.
\end{Thm}
{\it Proof.}
We provide a sketch of the proof only.
We apply Lieb's argument in \cite{Lieb}.

Since $S^{(3)}$ and $S_{\mathrm{tot}}^2$ are conserved,
we work in the $S^{(3)}=0$ subspace.
 By  putting $g_{xy}=0$ in Theorem \ref{Uni}, we know that  the ground
 state of $\hh$ in
 the  $S^{(3)}=0$
subspace is unique.
For each $U_0\ge 0$, let $H_{\mathrm{H}}(U_0)
=H_{\mathrm{H}}+\sum_{x\in \Lambda}U_0(n_x-1)^2
$. Since $\{U_{xy}\}$ is positive definite, so is $\{U_{xy}+2U_0
 \delta_{xy}\}$. Thus, the ground state of $\hh(U_0)$ in the $S^{(3)}=0$
 subspace is unique for all $U_0\ge 0$. By the continuity, the value of $S$ of
 the ground state of $\hh(U_0)$ in the $S^{(3)}=0$  subspace is independent of $U_0$.

Let $P=\prod_{x\in \Lambda}(n_{x\uparrow}-n_{x\downarrow})^2$.
Then it is known that 
\begin{align}
\| \{ \mathscr{W}U_0 \hh(U_0) \mathscr{W}^{-1}-h\}P \| \to 0 \ \ \ \mbox{as $U_0\to \infty$},\label{SW}
\end{align} 
where $h$ is the antiferromagnetic  Heisenberg model defined by
\begin{align}
h=\sum_{x, y\in \Lambda} J_{xy} (\mathbf{S}_x\cdot \mathbf{S}_y-\frac{1}{4})
\end{align} 
with $J_{xy}=2t_{xy}^2$ and $\mathscr{W}$ is the Schrieffer-Wolff  transformation.
By Marshall- Lieb-Mattis theorem \cite{LM}, 
the ground state of $hP$ is unique  and this state has total spin 
 $S=\frac{1}{2}\big| |A|-|B|\big|$. 
Since the ground state of $\mathscr{W} U_0\hh(U_0)\mathscr{W}^{-1}$ converges  to  that of $hP$,   
the value $S$ of the
 ground state of $\hh(U_0)$ must be identical to that of $hP$.
 $\Box$

\subsection{Proof of Theorem \ref{HH}}

In this proof, we work in the $S^{(3)}=0$ subspace, because $S^{(3)}$ and
$S_{\mathrm{tot}}^2$ are conserved.
Because the boson operators are unbounded, the proof has to be addressed
carefully. 

Our proof is an extension of  Lieb's argument in \cite{Lieb}.
For each $\theta\in [1, \infty)$, let $H_{\theta}$ be the Hamiltonian
$H$ with $\omega$ replaced by $\theta \omega$. Of course,
$H_{\theta=1}=H$.

\begin{lemm}\label{UniqueGS2}
The ground state of $H_{\theta}$ in the $S^{(3)}=0$ subspace is unique
 for all $\theta \ge 1$. 
\end{lemm} 
{\it Proof.}
By Theorem \ref{Uni}, it suffices to show that 
$
\{
U_{xy}-\frac{2}{\theta \omega}\sum_{z\in \Lambda}g_{xz} g_{yz}
\}_{x, y}
$ is positive definite for all $\theta \ge 1$.

First, we claim that  the matrix $
\{
\frac{2}{\omega}\sum_{z\in \Lambda}g_{xz} g_{yz}
\}_{x, y}
$ is positive semidefinite.
To see this,  
 let 
\begin{align}
M_{xy}=\frac{2}{\omega} \sum_{z\in \Lambda}g_{xz}g_{yz}. \label{MatrixM}
\end{align} 
Clearly, 
\begin{align}
\sum_{x,y\in \Lambda} \overline{\xi}_x \xi_yM_{xy}
=\frac{2}{\omega} \sum_{z\in \Lambda} \Bigg|
\sum_{x\in \Lambda}\xi_x g_{xz}
\Bigg|^2\ge 0
\end{align} 
for all $\{\xi_x\}\in \BbbC^{|\Lambda|}$.
 Hence, $\{M_{xy}\}$ is positive semidefinite.

Since $\{U_{\mathrm{eff}, xy}\}$ is positive definite, we have 
$
\sum_{x, y\in \Lambda}\overline{\xi}_x \xi_y 
U_{\mathrm{eff},xy}
>0
$ for all $\{\xi_x\}_{x\in \Lambda} \in \BbbC^{|\Lambda|}\backslash \{{\bf 0}\}$.
Therefore, we obtain 
\begin{align}
&\sum_{x,y\in \Lambda} \overline{\xi}_x \xi_y (U_{xy}-
\theta^{-1} M_{xy}
) \no
=&\sum_{x, y\in \Lambda} \overline{\xi}_x \xi_yU_{\mathrm{eff},xy}
+(1-\theta^{-1}) \sum_{z,y\in \Lambda} \overline{\xi}_x \xi_yM_{xy}>0
\end{align} 
for all $\{\xi_x\}_{x\in \Lambda} \in \BbbC^{|\Lambda|}\backslash \{{\bf 0}\}$.
Accordingly, $
\{
U_{xy}-\theta^{-1}M_{xy}
\}
$ is positive definite for all $\theta\ge 1$. $\Box$ 
\medskip\\

The Lang-Firsov transformation \cite{LF} is defined by $e^L$ with 
\begin{align}
L
=(\theta \omega)^{-1}\sum_{x, y\in \Lambda}g_{xy}n_x(b_y^*-b_y)
.\end{align}
  Set $H'_{\theta}=e^L H_{\theta}e^{-L}$.
We have
\begin{align}
H'_{\theta}=&\sum_{x, y\in \Lambda}\sum_{\sigma} t_{xy}e^{i\Phi_{xy}}
 c_{x\sigma}^*c_{y\sigma}
+
\theta \omega N_{\mathrm{b}}+\no
&+\sum_{x, y\in \Lambda}\Bigg(
U_{xy}-\frac{2}{\theta\omega}\sum_{z\in \Lambda} g_{xz}g_{yz}
\Bigg)(n_x-1)(n_y-1),
\end{align} 
where  $\Phi_{xy}
= -i (\theta \omega)^{-1}\sum_{z\in \Lambda} (g_{xz}-g_{yz}) (b_z^*-b_z)
$.

We rewrite  $H'_{\theta}$ as 
 $
H'_{\theta}=H_{\mathrm{H}}+\Delta_{\theta}+ \theta\omega N_{\mathrm{b}},
$ where  
\begin{align}
\Delta_{\theta}=\sum_{x, y\in \Lambda}\sum_{\sigma\in \{\uparrow, \downarrow\}} t_{xy}(e^{i\Phi_{xy}}-1)
 c_{x\sigma}^* c_{y\sigma}
-\sum_{x, y\in \Lambda} \theta^{-1}M_{xy} 
(n_x-1)(n_y-1),
\end{align} 
where $M_{xy}$ is given by (\ref{MatrixM})
\medskip

\begin{lemm}\label{lemm}
Let $K_{\theta}=H_{\mathrm{H}}+\theta \omega N_{\mathrm{b}}$. 
We have
 \begin{align}
\|\Delta_{\theta}(K_{\theta}-z)^{-1}\| \le C \theta^{-1}\bigg(
1+\frac{1+|z|}{|\mathrm{Im} z|}
\bigg)
\end{align} 
 for all $z\in \BbbC\backslash \BbbR$, where
$C$ is a positive constant independent of $\theta$ and $z$.
\end{lemm}
{\it Proof.}
Let \begin{align}
T=\sum_{x, y\in \Lambda}\sum_{\sigma\in \{\uparrow, \downarrow\}} t_{xy}(e^{i\Phi_{xy}}-1)
 c_{x\sigma}^* c_{y\sigma}.
\end{align}
  Since $\|(e^{iA}-1) \phi\| \le \|A\phi\|$ for any self-adjoint
operator $A$, we have
\begin{align}
\|T\phi\| \le C_1 \sum_{x, y\in \Lambda} \|\Phi_{x y} \phi\|, \
 \phi\in \D(N_{\mathrm{b}}) \label{Iq1}
\end{align}  
where $C_1$ is independent of $\theta$. Using the well-known bounds \footnote{
{\it Proof of the bounds.} Observe that 
\begin{align}
\|b_x \phi\|^2=\la \phi|b_x^*b_x\phi\ra \le
\la \phi|N_{\mathrm{b}}\phi\ra
     \le \|N_{\mathrm{b}} \phi\|^2.
\end{align} 
On the other hand, by the commutation relation $[b_x, b_x^*]=1$, we have
\begin{align}
\|b_x^* \phi\|^2=\|\phi\|^2 +\|b_x\phi\|^2 \le
 \|\phi\|^2+\|N_{\mathrm{b}} \phi\|^2.
\end{align} 
Since $\|N_{\mathrm{b}} \phi\| \le \|(N_{\mathrm{b}}+1) \phi\|$ and 
$ \|\phi\|^2+\|N_{\mathrm{b}} \phi\|^2 \le \|(N_{\mathrm{b}}+1)
	\phi\|^2$, 
we obtain the desired bounds.
}
:
$
\|b_x\phi\| \le \|(N_{\mathrm{b}}+1)\phi\|
$ and $\|
b_x^* \phi
\| \le \|(N_{\mathrm{b}}+1)\phi\|
$, we have 
\begin{align}
\|\Phi_{xy} \phi\| \le C_2\,  \theta^{-1} \|(N_{\mathrm{b}}+1) \phi\|, \label{Iq2}
\end{align} 
where $C_2$ is a positive constant independent of $\theta$.
Combining (\ref{Iq1}) and (\ref{Iq2}), we have 
\begin{align}
\|T\phi\|& \le C_3\,  \theta^{-1} \|(N_{\mathrm{b}}+1) \phi\|, \label{TInq}
\end{align} 
where $C_3$ is a positive constant independent of $\theta$.

Since 
\begin{align}
N_{\mathrm{b}}=(\omega \theta)^{-1}
 \{(K_{\theta}-z)-(H_{\mathrm{H}}-z)\}, 
\end{align} 
we have
\begin{align}
\|(N_{\mathrm{b}}+1) \phi\| \le (\theta \omega)^{-1} 
\Big\{
\|(K_{\theta}-z) \phi\|+(\|H_{\mathrm{H}}\|+1+|z|) \|\phi\|
\Big\}.
\end{align} 
Hence, 
\begin{align}
\|T\phi\| \le C_3 \omega^{-1}\theta^{-2} \Big\{
\|(K_{\theta}-z) \phi\|+(\|H_{\mathrm{H}}\|+1+|z|) \|\phi\|
\Big\}.
\end{align} 
Because  $\|
\sum_{x, y\in \Lambda} \theta^{-1}M_{xy} 
(n_x-1)(n_y-1) \| \le C_4\,  \theta^{-1}
$ with $C_4$, a positive constant independent of $\theta$, we have 
\begin{align}
\|\Delta_{\theta} \phi\| \le \theta^{-1} C  \Big\{
\|(K_{\theta}-z) \phi\|+(\|H_{\mathrm{H}}\|+1+|z|) \|\phi\|
\Big\}.
\end{align} 
Using  $\|(K_{\theta}-z)^{-1}\| \le |\mathrm{Im} z|^{-1}$,
 we obtain the desired bound.
$\Box$
\medskip\\

\begin{lemm}\label{LemConv}
For all $z\in \BbbC\backslash\BbbR$, we have
\begin{align}
\lim_{\theta\to \infty}
\|
(H_{\mathrm{H}}-z)^{-1}\otimes P_{\Omega}-(H'_{\theta}-z)^{-1}
\|= 0, \label{Conv}
\end{align}
 where $P_{\Omega}=|\Omega\ra\la \Omega|$ with $\Omega$,  the bosonic
Fock vacuum. 
\end{lemm} 
{\it Proof.}
By Lemma \ref{lemm} and the fact $\|(H'_{\theta}-z)^{-1 }\| \le |\mathrm{Im} z|^{-1}$, we have 
\begin{align}
\|
(H'_{\theta}-z)^{-1}-(K_{\theta}-z)^{-1}
\| 
&=\|(H'_{\theta}-z)^{-1} \Delta_{\theta} (K_{\theta}-z)^{-1}\|
\no
&\le  C \theta^{-1}|\mathrm{Im }z|^{-1} 
\bigg(
1+\frac{1+|z|}{|\mathrm{Im}z|}
\bigg)
\to 0  \label{Resolvent1}
\end{align}
 as $\theta \to \infty$ for every $z\in \BbbC\backslash \BbbR$.
 
On the other hand, we  obtain  that 
\begin{align}
\|
(K_{\theta}-z)^{-1}-(H_{\mathrm{H}}-z)^{-1}\otimes P_{\mathrm{\Omega}}
\| \to 0\label{Resolvent2}
\end{align} 
as $\theta\to \infty$.
To see this, 
we decompose the $S^{(3)}=0$ subspace as 
\begin{align}
\mathcal{H}_{M=0} =\bigoplus_{n=0}^{\infty} \mathcal{K}_n,\quad
 \mathcal{K}_n
=\mathcal{H}_{M=0} \cap \ker(N_{\mathrm{b}}-n).\label{BosonD}
\end{align} 
$\mathcal{K}_n$ is called the $n$ phonon subspace.
Corresponding to (\ref{BosonD}), we have
\begin{align}
K_{\theta}=\bigoplus_{n=0}^{\infty} \big(H_{\mathrm{H}}+\theta \omega n \big),
\end{align} 
which implies
\begin{align}
(K_{\theta}-z)^{-1}= \bigoplus_{n=0}^{\infty} \big(H_{\mathrm{H}}+\theta
 \omega n-z\big)^{-1}
\end{align} 
for all $z\in \BbbC\backslash \BbbR$.
Let $e$ be the lowest energy of $H_{\mathrm{H}}$ in the $S^{(3)}=0$ subspace.
If $\theta$ is large enough such that $e +\theta \omega-|\mathrm{Re}z|>0$,
we obtain
\begin{align}
\big\|
\big(
H_{\mathrm{H}}+\theta \omega n-z
\big)^{-1}
\big\| \le \big(e+\theta \omega-|\mathrm{Re}z|\big)^{-1}
\end{align} 
for all $n\ge 1$. Therefore, 
\begin{align}
\big\|
(K_{\theta}-z)^{-1}-(H_{\mathrm{H}}-z)^{-1} \otimes P_{\Omega}
\big\|
=&\sup_{n\ge 1} \big\|
\big(
H_{\mathrm{H}}+ \theta \omega n-z 
\big)^{-1}
\big\|\no
\le&  \big(e+\theta \omega-|\mathrm{Re}z|\big)^{-1}
\to 0
\end{align} 
as $\theta \to \infty$ for $z\in \BbbC\backslash \BbbR$.

By (\ref{Resolvent1}) and (\ref{Resolvent2}), we obtain (\ref{Conv}).
 $\Box$

\begin{lemm}\label{Conti}
Let $E_0(\theta)$ and $E_1(\theta)$ be the ground state energy and the
first excited energy of $H'_{\theta}$ in the $S^{(3)}=0$ subspace,
 respectively. In addition, let $E_0$ and $E_1$ be the ground state
 energy and the first excited energy of $H_{\mathrm{H}}$ in the
 $S^{(3)}=0$
subspace, respectively.
\begin{itemize}
\item[{\rm (i)}]  $E_0(\theta)$ converges to $E_0$,
 and $E_1(\theta)$ converges
to $E_1$ as $\theta\to
	     \infty$, respectively.
\item[{\rm (ii)}] $E_0(\theta)$ and $E_1(\theta)$ are continuous in $\theta$.
\end{itemize} 
\end{lemm} 
{\it Proof.}
(i)  follows from Lemma \ref{LemConv}.

(ii) Note that $\D(H_{\theta})=\D(N_{\mathrm{b}})$ for all $\theta$.
In addition, $H_{\theta} \phi$ is a vector-valued analytic function of
$\theta$ for all $\phi\in \D(N_{\mathrm{b}})$. Thus, $H_{\theta}$ is an analytic family of type (A)
\cite{ReSi4}
 in a neighborhood  of $[1,\infty) \subset \BbbC$.
By \cite[Theorem XII. 13]{ReSi4}, $E_0(\theta)$ and $E_1(\theta)$ are
analytic, in particular, continuous in $\theta$.
 $\Box$

\begin{lemm}\label{Gap}
Set $
\delta:=\inf_{\theta \ge 1}|E_1(\theta)-E_0(\theta)|
$.
We have $\delta>0$.

\end{lemm} 
{\it Proof}.
We claim that  
 $E_0(\theta)\neq E_1(\theta)$ for all $\theta \ge 1$. Indeed, 
assume that there exists a $\theta_0\ge 1$ such
that $E_0(\theta_0)=E_1(\theta_0)$. Then the uniqueness of  the ground
states is  broken  at $\theta=\theta_0$, which contradicts  
with Lemma \ref{UniqueGS2}.

Because  $E_1-E_0>0$, we get $ \delta
>0
$ by Lemma \ref{Conti}. $\Box$
\medskip\\

Let $\psi_{\theta}$ be the ground state of $H'_{\theta}$ and let $\psi$ be the
ground state of $H_{\mathrm{H}}$ in the $S^{(3)}=0$ subspace.
Remark that these are unique ground states of $H_{\theta}'$ and
$H_{\mathrm{H}}$ by Lemma \ref{UniqueGS2}.

\begin{lemm}\label{ConstS}
Let $S_{\theta}$ be the total spin of  $\psi_{\theta}$:
 $S_{\mathrm{tot}}^2\psi_{\theta}=S_{\theta}(S_{\theta}+1)
 \psi_{\theta}$.
The value of $S_{\theta}$ is independent of $\theta \ge 1$.
\end{lemm} 
{\it Proof.} 
First, we claim  that $\psi_{\theta}$ is continuous in $\theta$, namely,
\begin{align}
\lim_{\theta'\to \theta}\|\psi_{\theta}-\psi_{\theta'}\|=0.\label{ResCont}
\end{align} 
 Indeed, since $H_{\theta}\phi$ is continuous in $\theta$ for all
$\phi\in \D(N_{\mathrm{b}})$, $(H_{\theta}-z)^{-1} \phi $ is continuous
in $\theta$  for all $\phi\in \D(N_{\mathrm{b}})$ by \cite[Theorem VIII 25
]{ReSi}. 
Here, we used the fact that $\D(H_{\theta})=\D(N_{\mathrm{b}})$ for all
$\theta \ge 1$.
Thus, applying \cite[Theorem VIII 24]{ReSi}, we conclude (\ref{ResCont}).

Since $S_{\mathrm{tot}}^2$ is bounded, we have 
\begin{align}
\Big|
S_{\theta}(S_{\theta}+1)-S_{\theta'}(S_{\theta'}+1)
\Big|
&=
\big|
\|S_{\mathrm{tot}}^2\psi_{\theta}\|-\|S_{\mathrm{tot}}^2 \psi_{\theta'}\|
\big|\no
&\le \|S_{\mathrm{tot}}^2 (\psi_{\theta}-\psi_{\theta'})\| \no 
 &\le \|S_{\mathrm{tot}}^2\| \| \psi_{\theta}-\psi_{\theta'}\| \to 0
\end{align} 
as $\theta\to \theta'$. Thus, $S_{\theta}$ is continuous in $\theta$.
 On the other hand, because $S_{\theta}$ takes discrete values, it must
 be independent of $\theta \ge 1$. $\Box$
\medskip\\

\begin{flushleft}
{\it Completion of proof of Theorem \ref{HH}}
\end{flushleft} 
First, we remark the following formula:
\begin{align}
|\psi_{\theta}\ra\la \psi_{\theta}|&=\frac{i}{2\pi}
 \oint_{|E-E_0|=\delta/2}(H'_{\theta}-E)^{-1}dE\ \ \mbox{for all
 $\theta\ge 1$}, \label{Re1}\\
|\psi\ra\la \psi| \otimes P_{\Omega}&=\frac{i}{2\pi}
 \oint_{|E-E_0|=\delta/2}(H_{\mathrm{H}}-E)^{-1} \otimes P_{\Omega}dE, \label{Re2} 
\end{align} 
where $\delta$ is given in Lemma \ref{Gap}.
By (\ref{Conv}), (\ref{Re1}) and (\ref{Re2}),
we have $
\|\psi_{\theta}-\psi\otimes \Omega\|\to 0
$ as $\theta\to \infty$. 
Recall  that the value of $S$ of $\psi_{\theta}$ must be independent of $\theta$ 
by Lemma \ref{ConstS}.
Since the ground state $\psi\otimes \Omega$ has total  spin
$S=\frac{1}{2}\big||B|-|A|\big|$ by Theorem \ref{Hubbard}, 
 so does $\psi_{\theta}$ due to  the continuity. 
To see this, suppose that $\psi_{\theta}$ has total spin $S'$ for all $\theta
\ge 1$. 
By Lemma \ref{ConstS}, $S'$ is independent of $\theta$.
We have 
\begin{align}
|S(S+1)-S'(S'+1)| 
&\le \big\|
S_{\mathrm{tot}}^2(\psi_{\theta}-\psi\otimes \Omega)
\big\|\no
&\le \|S_{\mathrm{tot}}^2\|\|\psi_{\theta}-\psi\otimes \Omega\|\no
& \to 0
\end{align} 
as $\theta\to \infty$. Hence, $S'=S$.
$\Box$
\medskip\\

\subsection{Proof of Theorem \ref{LRO}}

We follow \cite{Shen}. By Theorem \ref{Uni},  we obtain that 
\begin{align*}
m(0) &= |\Lambda|^{-1} \sum_{x,y}\la S_x^{(+)} S_y^{(-)} \ra\no
&\le |\Lambda|^{-1}\sum_{x,y} \gamma(x) \gamma(y)\la S_x^{(+)} S_y^{(-)}
\ra\no
&=m(Q).
\end{align*} 
Since $m(0)=O(\Lambda)$ by Theorem \ref{HH}, we conclude the assertions in 
Theorem \ref{LRO}. $\Box$
\medskip

\subsection{Proof of Theorem \ref{Ab}}

We provide a sketch only. We apply Kubo-Kishi argument \cite{KK}, which
originates from \cite{DLS}, see also \cite{KLS}.
For each ${\boldsymbol h}=\{h_x\}_{x\in \Lambda}\in \BbbR^{\Lambda}$, 
let $H'({\boldsymbol  h})$ be the Hamiltonian $H'_{\theta=1}$ with
$\Ue=\frac{1}{2}\sum_{x, y\in \Lambda} U_{\mathrm{eff}, xy} (n_{x\uparrow}-n_{x\downarrow})(n_{y\uparrow}-n_{y\downarrow})$ replaced by 
$
\Ue({\boldsymbol h})=\frac{1}{2}\sum_{x, y\in \Lambda} U_{\mathrm{eff}, xy}
(n_{x\uparrow}-n_{x\downarrow}+h_x)(n_{y\uparrow}-n_{y\downarrow}+h_y)$.
Clearly, we have $H'(\mb{0})=H'_{\theta=1}$.
 We denote by $\mathcal{H}$ the
$S^{(3)}=0$
 subspace. Let $Z_{\beta}({\boldsymbol h})=\Tr_{\mathcal{H}}[e^{-\beta
 H'({\boldsymbol h})}]$. Then we can show that  $Z_{\beta}({\boldsymbol h}) \le
Z_{\beta}(\mb{0})$, see \cite{Miyao} for details. This implies that  $E(\mb{0}) \le
E({\boldsymbol h}) $, where $E({\boldsymbol h})$ is the ground state energy of
$H'({\boldsymbol h})$ in the $S^{(3)}=0$ subspace. Thus, we get $d^2E(\lambda
{\boldsymbol h})/d\lambda^2|_{\lambda=0} \ge 0$, which implies  Theorem \ref{Ab}. $\Box$ 

\begin{flushleft}
{\bf Acknowledgments.}
I would like to thank  the anonymous referee for valuable  comments.
This work was partially supported by KAKENHI (18K0331508) and   KAKENHI
 (16H03942). 
\end{flushleft}

\end{document}